\documentclass[overfull,debug]{epl}

\usepackage{graphics}
\usepackage{epsfig}
\usepackage{amsmath}
\usepackage{psfrag}

\title{Super-Arrhenius dynamics for sub-critical crack growth in disordered brittle media}
\shorttitle{Super-Arrhenius dynamics for sub-critical crack
growth}
\author{P.-P. Cortet \and L. Vanel, \and S. Ciliberto.}
\shortauthor{P.-P.Cortet \etal}

\institute{
  Laboratoire de Physique, CNRS UMR 5672,
  Ecole Normale Sup\'erieure de Lyon,
  46 all\'ee d'Italie,
  69364 Lyon Cedex 07, France \\
}


\begin{document}

\maketitle

\begin{abstract}
Taking into account stress fluctuations due to thermal noise, we
study thermally activated irreversible crack growth in disordered
media. The influence of material disorder on sub-critical growth
of a single crack in two-dimensional brittle elastic material is
described through the introduction of a rupture threshold
distribution. We derive analytical predictions for crack growth
velocity and material lifetime in agreement with direct numerical
calculations. It is claimed that crack growth process is inhibited
by disorder: velocity decreases and lifetime increases with
disorder. More precisely, lifetime is shown to follow a
super-Arrhenius law, with an effective temperature
$\theta-\theta_d$, where $\theta$ is related to the
thermodynamical temperature and $\theta_d$ to the disorder
variance.
\end{abstract}

\section{Introduction}
Sub-critical rupture occurs when a material is submitted to a
stress lower than a critical threshold. It is known since the
early work of Zhurkov or Brenner \cite{Zhurkov, Brenner} that
sub-critical rupture can be thermally activated. However,
prediction of the lifetime, i.e. the time it takes for a sample to
break under a given load, has proved difficult when the material
is heterogeneous. In addition, although there have been several
theoretical attempts to predict the lifetime for homogeneous
\cite{Golubovic, Pomeau, Buchel} and sometimes disordered
\cite{Arndt, Roux} elastic materials, there have been very few
theoretical studies of the actual rupture dynamics in
heterogeneous media when thermal activation is the driving
mechanism \cite{Ciliberto, Politi, Sornette}. Recently, this
dynamics has been modelled for a single macroscopic crack in
two-dimensional homogeneous elastic media taking into account
stress fluctuations \cite{Santucci1}. This model has been
successfully faced with experiments of crack growth in fax paper
sheets even though this material is actually heterogeneous
\cite{Santucci2}. In the present letter, we will take into account
heterogeneity by introducing, in the thermal activation rupture
model, disorder in the rupture thresholds. We will show that
disorder slows down macroscopic crack growth contrary to previous
results \cite{Ciliberto, Politi} that show the rupture dynamics to
be accelerated by disorder in the case of diffuse damage.

\section{Model for sub-critical crack growth in homogeneous brittle media}
In ref. \cite{Santucci1}, the model that describes sub-critical
crack growth in homogeneous brittle media assumes that thermal
fluctuations induce local stress fluctuations with time in the
elastic material according to a gaussian distribution:
\begin{equation}\label{stressdist}
  p(\sigma) = \frac{1}{\sqrt {2\pi \theta}}
  \exp \left[ -\frac{\left(\sigma-\overline{\sigma} \right)^2}{2\theta}
  \right]\text{,}
\end{equation}
with $\theta$ the temperature in square stress unit:
$\theta=Yk_BT/V$, where Y is the Young modulus, $k_B$ the
Boltzmann constant and $T$ the thermodynamical temperature. In eq.
(\ref{stressdist}), $\overline{\sigma}$ is the equilibrium value
of the stress $\sigma$ fluctuating in a volume $V$. We assume the
volume $V$ to break if the fluctuating local stress $\sigma$
becomes larger than a critical threshold $\sigma_c$. In the
two-dimensional case where a single crack is loaded in mode $1$
(uniaxial stress perpendicular to the crack direction), stress
concentration makes almost certain that breaking occurs at the
crack tip. Actually, at the crack tip, linear elasticity predicts
a divergence of the stress. However, the average elastic stress
$\sigma_m$ on a volume $V\sim\lambda^3$ at the crack tip does not
diverge and can be estimated as:
\begin{equation}\label{ctstress}
\sigma_m(\ell) =
2\frac{\sigma_e\sqrt{\ell}}{\sqrt{2\pi\lambda}}\text{,}
\end{equation}
where $\ell$ is the crack length and $\sigma_e$ the applied
stress. The local stress on the crack tip volume $V$ is submitted
to time fluctuations according to eq. (\ref{stressdist}) with
$\overline{\sigma}=\sigma_m(\ell)$.

The cumulative probability that the instantaneous crack tip stress
is larger than the threshold $\sigma_c$ is $P(\sigma_c) =
\int_{\sigma_c} ^{+\infty} {p\left(x\right) \upd x}$. Then, the
average time $\langle t_{w}\rangle$ needed by the crack tip
material to break on the mesoscopic scale $\lambda$ is directly
proportional to the inverse of this cumulative probability:
$\langle t_{w}\rangle=t_0/P(\sigma_c)$, where $t_0$ is a
characteristic time for stress fluctuations. The statistically
averaged inverse crack velocity, $dt/d\ell$, is constructed as the
ratio of the average time $\langle t_{w}\rangle$ over the growth
length due to rupture at each tip of the crack, $dt/d\ell=\langle
t_{w}\rangle/2\lambda$. Then,

\begin{equation}\label{velocity1}
 \frac{\upd t(\ell) }{\upd \ell} =\left\{\begin{array}{ll}
\left[v_0\,\text{erfc}\left(\frac{\sigma_c-\sigma_m(\ell)}
{\sqrt{2\theta}}\right)\right]^{-1} & \text{if $\sigma_c-\sigma_m(\ell)\geq 0$}\\
v_0^{-1} & \text{if $\sigma_c-\sigma_m(\ell)<
0$}\end{array}\right.
=\frac{1}{v_h(\sigma_m(\ell),\sigma_c,\theta)} \text{,}
\end{equation}
with $v_0=\lambda/t_0$. For critical crack growth, i.e. when
$\sigma_m(\ell)>\sigma_c$, the crack simply propagates at its
characteristic dynamical velocity: $dt/d\ell=1/v_0$.

This model was successfully used to describe the growth of a
single crack in a fibrous material such as paper \cite{Santucci2}.
In that case, the scale $\lambda$ corresponds to the typical paper
fiber diameter. However, paper is heterogeneous and distributions
in fiber size, position or orientation introduce some disorder
which has not been taken into account in this model.

\section{Model extension for disordered media}
Many brittle materials are heterogeneous at a mesoscopic scale.
Crack front roughness is one evidence for this heterogeneity. To
introduce some disorder in the model, we assume the rupture
threshold of the material, $\sigma_c$, to be distributed at the
scale $\lambda$ with a gaussian distribution:

\begin{equation}\label{distribution}
  p_{th}(\sigma_c) = \frac{1}{\sqrt {2\pi \theta_d}}
  \exp \left[ -\frac{\left(\sigma_c-\overline{\sigma_c}\right)^2}{2\theta_d}
  \right]\, \text{for $\sigma_c\geq0$,}
\end{equation}
with a variance verifying $\sqrt{\theta_d}\ll\overline{\sigma_c}$
so that the distribution can be considered normalized. For sake of
simplicity, we will consider the crack to grow on a straight line
and introduce disorder effects only through the threshold
distribution. Then, it is clear that, during its growth, the crack
is faced, in statistical average, with the whole intrinsic
threshold distribution of the material. The relevance of this
simplification will be discussed in the last section.

Using the threshold distribution of eq. (\ref{distribution}) at
the crack tip, a new expression for the statistical average of the
inverse crack velocity can be written:

\begin{equation}\label{velocity2}
 \frac{\upd t}{\upd \ell} = \int_{0}^{+\infty}
\frac{p_{th}(\sigma_c)\upd
\sigma_c}{v_h(\sigma_m(\ell),\sigma_c,\theta)}\text{,}
\end{equation}
where $v_h$ is the crack velocity for the homogeneous case. Here,
the inverse velocity, $1/v_h$, is weighted by the threshold
distribution because the waiting time for the crack to grow by a
fixed step $\lambda$ is the statistical variable. Assuming
$\overline{\sigma_c}-\sigma_m\gg \sqrt{\theta}+\sqrt{\theta_d}$,
the integrand in eq. (\ref{velocity2}) takes significative values
only for a range of $\sigma_c$ such that: (i)
$\sigma_c-\sigma_m\geq 0$ and the integral can be truncated from
$\sigma_m$ to $+\infty$; (ii) the energy barrier
$(\sigma_c-\sigma_m)^2/2\theta$ can be considered far larger than
one so that an asymptotic development of the complementary error
function can be introduced in eq. (\ref{velocity2}). After
rearrangement, we obtain:
\begin{equation}\label{velocity3}
\frac{\upd t}{\upd \ell} \simeq
\frac{e^{\frac{(\overline{\sigma_c} -\sigma_m)^2}
{2(\theta-\theta_d)}}}{2v_0\,\sqrt{\theta\theta_d}}\int_{\sigma_m}^{+\infty}
(\sigma_c-\sigma_m)e^{-\frac{\theta-\theta_d}{2\theta\theta_d}
\left(\sigma_c-\widetilde{\sigma_c}\right)^2}\upd \sigma_c
\text{,}
\end{equation}
with
\begin{equation}\label{sigmac}
 \widetilde{\sigma_c}= \frac{\theta\,\overline{\sigma_c}-\theta_d
 \,\sigma_m}{\theta-\theta_d}.
\end{equation}
Then, we can estimate the integral:
\begin{eqnarray}\label{int1}
\int_{\sigma_m}^{+\infty} (\sigma_c-\sigma_m)
e^{-\frac{\theta-\theta_d}{2\theta\theta_d}
\left(\sigma_c-\widetilde{\sigma_c}\right)^2}\upd \sigma_c =
\int_{\sigma_m}^{+\infty}
(\sigma_c-\widetilde{\sigma_c}+\widetilde{\sigma_c}-\sigma_m)
e^{-\frac{\theta-\theta_d}{2\theta\theta_d}
\left(\sigma_c-\widetilde{\sigma_c}\right)^2}\upd \sigma_c\nonumber\\
=\left[-\frac{\theta\theta_d}{\theta-\theta_d}e^{-\frac{\theta-\theta_d}{2\theta\theta_d}
\left(\sigma_c-\widetilde{\sigma_c}\right)^2}\right]_{\sigma_m}^{+\infty}+
(\widetilde{\sigma_c}-\sigma_m)\int_{\sigma_m}^{+\infty}
e^{-\frac{\theta-\theta_d}{2\theta\theta_d}
\left(\sigma_c-\widetilde{\sigma_c}\right)^2}\upd \sigma_c.
\end{eqnarray}

Here, it is important noticing that both terms of eq. (\ref{int1})
are infinite when $\theta<\theta_d$. So, it is clear that, in the
case where $\theta<\theta_d$, the inverse crack velocity and the
rupture time will become infinite. On the other hand, considering
the case where $\theta>\theta_d$, we get:
\begin{equation}\label{velocity4}
\frac{\upd t}{\upd \ell} \simeq \frac{\theta}{\theta-\theta_d}
\frac{1}{v_0}
\sqrt{\frac{\pi}{2(\theta-\theta_d)}}(\overline{\sigma_c}-\sigma_m)\exp
\left[\frac{\left(\overline{\sigma_c}
-\sigma_m\right)^2}{2(\theta-\theta_d)}\right].
\end{equation}

Eq. (\ref{velocity4}) corresponds, with an additional prefactor
$\theta/(\theta-\theta_d)$, to the asymptotic development of eq.
(\ref{velocity1}) for the homogeneous case introducing an
effective temperature $\theta_{\text{eff}}=\theta-\theta_d$. Then,
eq. (\ref{velocity4}) can be seen as an approximation of:

\begin{equation}\label{velocity5}
\frac{\upd t}{\upd \ell} \simeq \frac{\theta}{\theta-\theta_d}
\frac{1}{v_{h}(\sigma_m,\overline{\sigma_c}, \theta-\theta_d)}\,
\text{, when $\theta>\theta_d$}.
\end{equation}

The conclusion of this part is that crack velocity is lowered by
the introduction of disorder in the model material, mainly through
the appearance of a lowered effective temperature
$\theta_{\text{eff}}=\theta-\theta_d$ in place of the
thermodynamical temperature $\theta$. The variance $\theta_d$ can
then be seen as a temperature of disorder.

\section{Direct numerical analysis of the model}
\begin{figure}[h]
    \psfrag{A}[c]{(a)}
    \psfrag{B}[c]{(b)}
    \psfrag{X}[c]{$\ell$}
    \psfrag{Y}[c]{$\,(\theta-\theta_d)/\theta\left(\frac{dt}{d\ell}\right)_{\text{num}}$}
    \psfrag{Z}[l][][0.7]{Fit by $y=1/v_{h}(\sigma_m(\ell),\overline{\sigma_c},\theta_{\text{eff}})$}
    \psfrag{H}[l][][0.7]{with $\theta_{\text{eff}}=0.0040$}
    \psfrag{W}[l][][0.7]{Numerical rhs of eq.
    (\ref{velocity2})}
    \psfrag{U}[c]{$\theta-\theta_d$}
    \psfrag{V}[c]{$\theta_{\text{eff}}$}
    \centerline{
    \includegraphics[width=6cm]{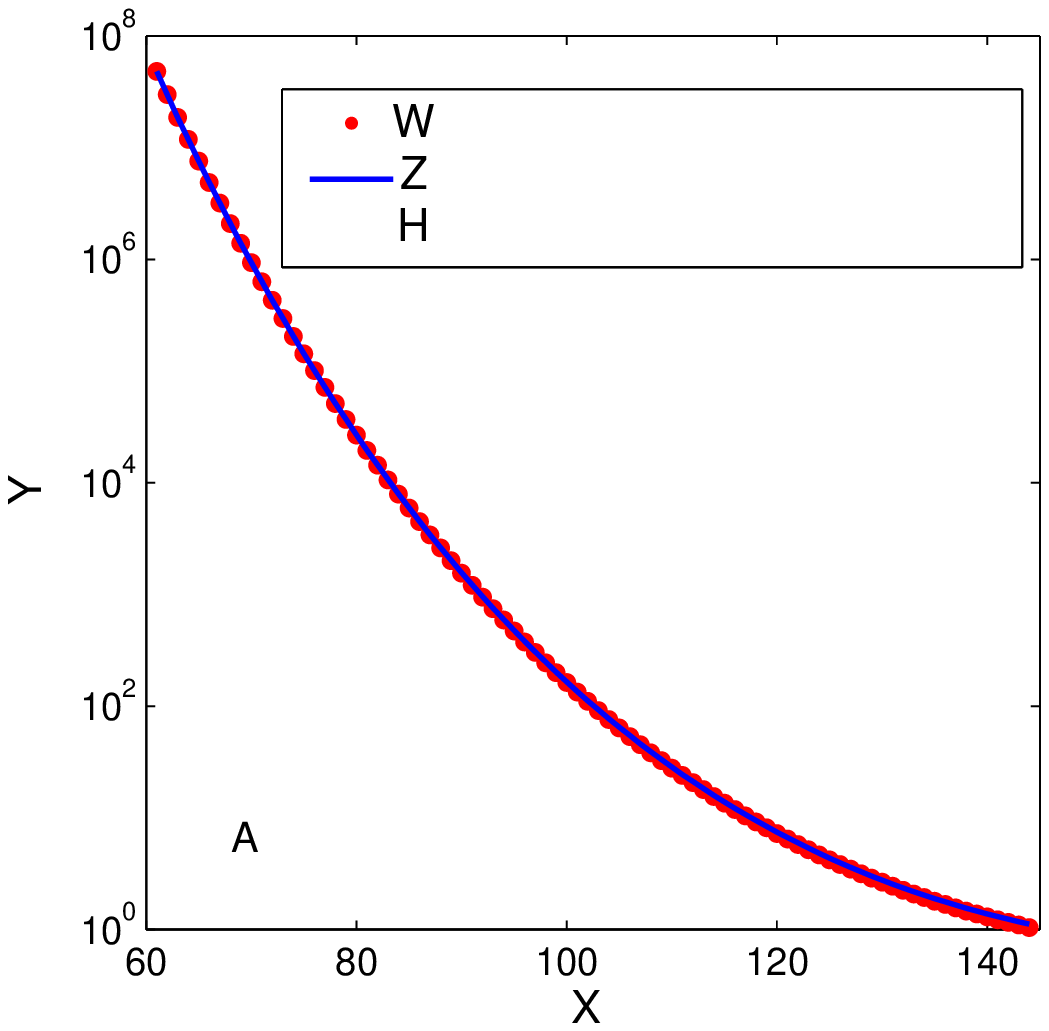}
    \includegraphics[width=5.9cm]{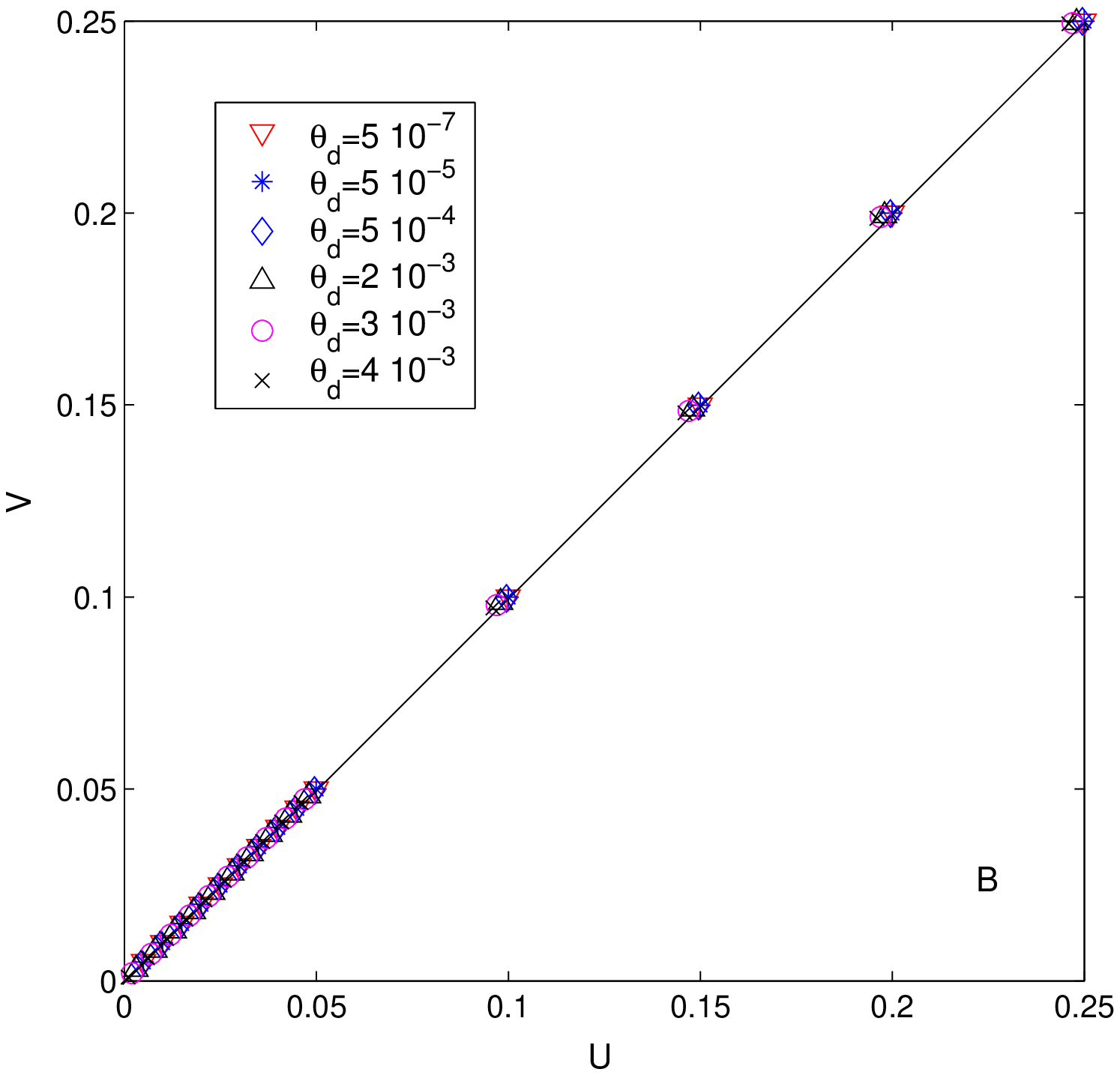}}
    \caption{On the left: Semi-log plot of numerical
$(\theta-\theta_d)/\theta\,(dt/d\ell)_{\text{num}}$ as a function
of $\ell$ ($v_0=1$, $\lambda=1$, $\overline{\sigma_c}=1$,
$\sigma_e=0.10$, $\theta=0.005$, $\theta_d=0.001$) and its fit by
$y=1/v_{h}(\sigma_m(\ell),\overline{\sigma_c},\theta_{\text{eff}})$;
on the right: $\theta_{\text{eff}}$ obtained from the fits as a
function of $\theta-\theta_d$, with $\theta$ and $\theta_d$ both
varying ($0.005\leq\theta\leq 0.25$ and
$5\,10^{-7}\leq\theta_d\leq 0.004$).} \label{vgraph}
\end{figure}

In order to check the validity of eq. (\ref{velocity5}), we study
numerically the model starting from eq. (\ref{velocity2}). In fig.
\ref{vgraph}a), we plot
$(\theta-\theta_d)/\theta\,(dt/d\ell)_{\text{num}}$, obtained by
numerical calculation of eq. (\ref{velocity2})'s right hand side,
as a function of the crack length for a set of particular values
for the model parameters. The analytical law,
$y=1/v_{h}(\sigma_m(\ell),\overline{\sigma_c},\theta_{\text{eff}})$,
fits very well the numerical data using $\theta_{\text{eff}}$ as a
fitting parameter.

It is clear as one can see in fig. \ref{vgraph}b) that the
effective temperature $\theta_{\text{eff}}$ obtained from the
numerical data fits corresponds very well to the one predicted
analytically, $\theta_{\text{eff}}=\theta-\theta_d$, when $\theta$
and $\theta_d$ are both varying. However, the coincidence of the
numerical $\theta_{\text{eff}}$ with $\theta-\theta_d$ is not
perfect. Numerical $\theta_{\text{eff}}$ is actually perfectly
described by $\theta_{\text{eff}}=a(\theta_d)\,(\theta-\theta_d)$
with $a$ ranging from $1$ to $1.014$ when $\theta_d$ increases
from $0$ to $0.004$ ($\overline{\sigma_c}=1$). These numerical
results confirm the good quality of eq. (\ref{velocity5}) even if
it is the result of approximate calculations.

\section{Rupture time}
We will now go a step further in the analysis of our model by
studying the rupture time dependence with the model parameters.
When there is no disorder, rupture time is defined as the time for
a crack of initial length $\ell_i$ to grow until it reaches a
critical length $\ell_c$ such that $\sigma_m(\ell_c)=\sigma_c$. In
paper \cite{Santucci1}, we got an approximate expression of the
mean rupture time for homogeneous systems assuming some
approximations in eq. (\ref{velocity1}):

\begin{equation}\label{lifetime}
\tau \simeq \frac{\sqrt{2 \pi \theta} \ell_i}{v_0 \sigma_i}
e^{\frac{\left(\sigma_c-\sigma_i\right)^2}{2\theta}}=\tau_o\,
e^{\frac{\left(\sigma_c-\sigma_i\right)^2}{2\theta}}\,\text{,
where $\sigma_i=\sigma_m(\ell_i)$.}
\end{equation}

This law has been tested experimentally on the slow growth of a
single crack in fax paper sheets \cite{Santucci2}. It is shown
that statistically averaged crack growth curves are in good
agreement with the model predictions as well as the lifetime
dependence on applied stress, initial crack length and Young
modulus.

Backing on the calculations of paper \cite{Santucci1}, we easily
get an analytical expression for the mean lifetime in the
heterogeneous case from eq. (\ref{velocity5}):

\begin{equation}\label{lifetime2}
\tau \simeq \frac{\theta}{\sqrt{\theta-\theta_d}}\frac{\sqrt{2
\pi} \ell_i}{v_0 \sigma_i}
e^{\frac{\left(\overline{\sigma_c}-\sigma_i\right)^2}{2(\theta-\theta_d)}}
=\tau_o^{des}\,e^{\frac{\left(\overline{\sigma_c}-\sigma_i\right)^2}
{2(\theta-\theta_d)}}\,\text{, when $\theta>\theta_d$.}
\end{equation}

\begin{figure}
    \psfrag{A}[c]{(a)}
    \psfrag{B}[c]{(b)}
    \psfrag{X}[c]{$\frac{(\overline{\sigma_c}-\sigma_i)^2}{2(\theta-\theta_d)}$}
    \psfrag{Y}[c]{$\text{log}(\tau/\tau_o^{des})$}
    \psfrag{U}[c]{$\frac{(\overline{\sigma_c}-\sigma_i)^2}{2\theta}$}
    \psfrag{V}[c]{$\text{log}(\tau/\tau_o)$}
    \centerline{
    \includegraphics[width=6.08cm]{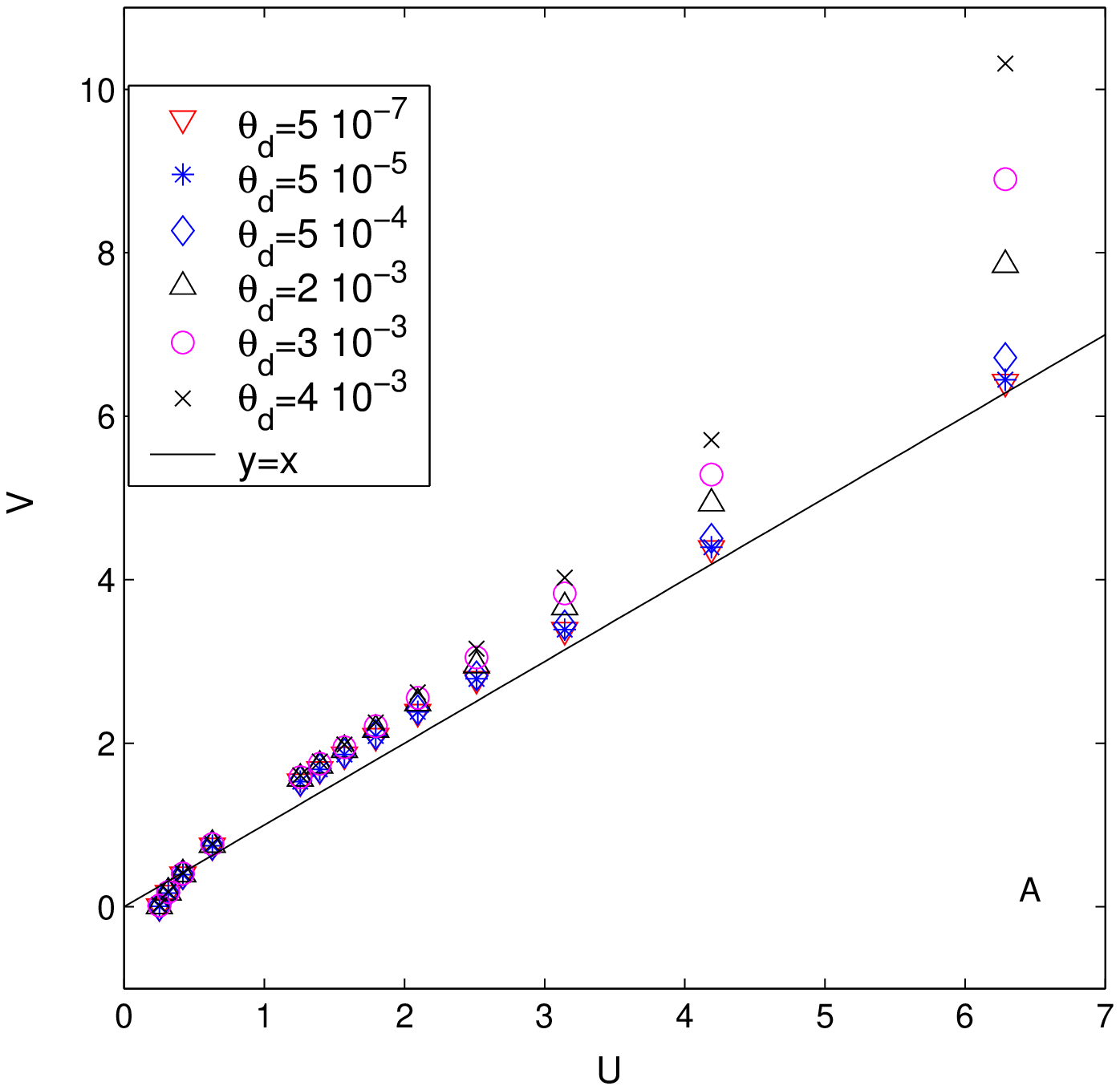}
    \includegraphics[width=6cm]{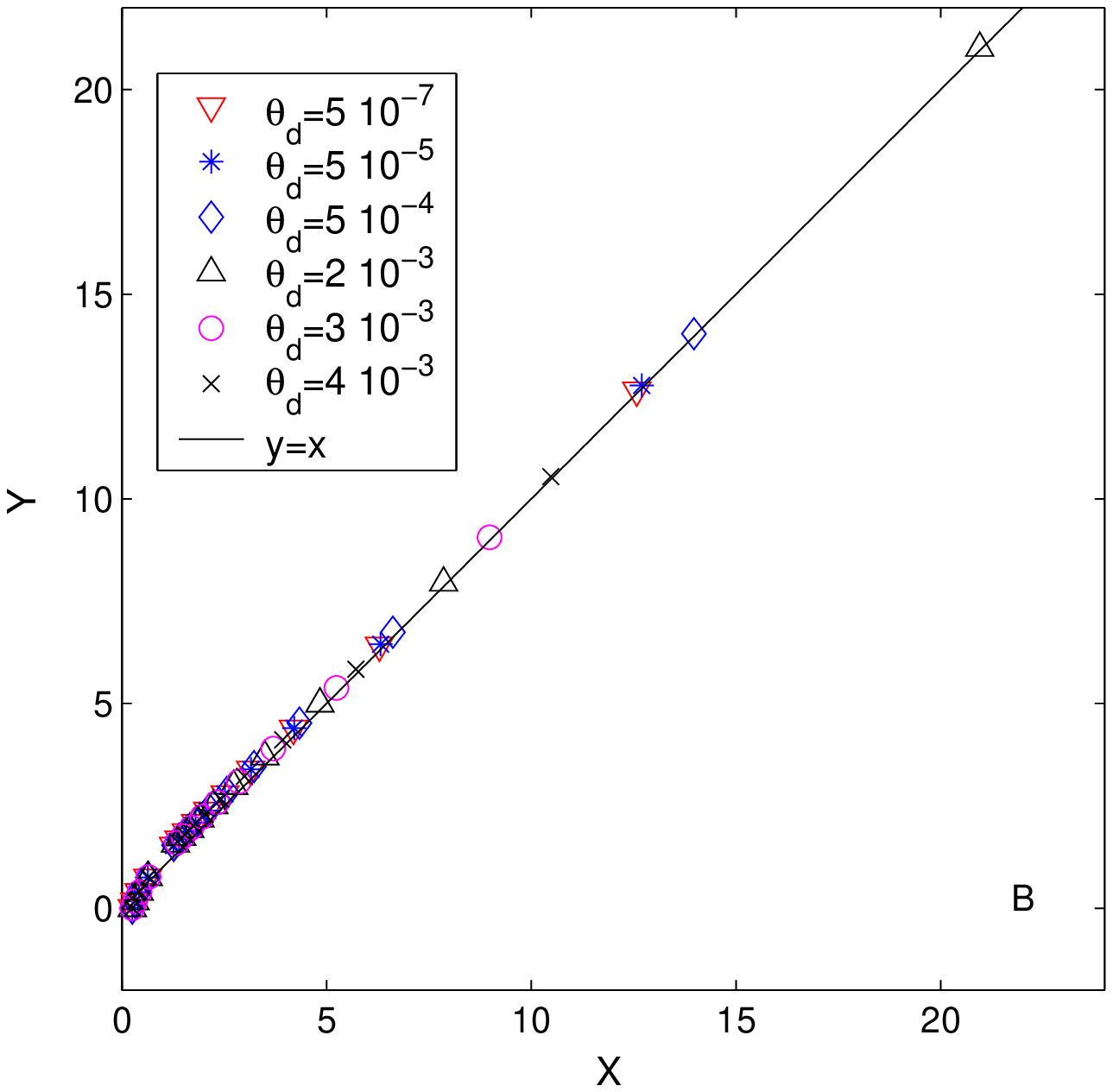}}
    \caption{On the left: logarithm of the rupture
time as a function of the energy barrier
$(\overline{\sigma_c}-\sigma_i)^2/2\theta$ for many different
growth conditions ($0.005\leq\theta\leq 0.25$ and
$5\,10^{-7}\leq\theta_d\leq 0.004$); on the right: logarithm of
the rupture time as a function of
$(\overline{\sigma_c}-\sigma_i)^2/2(\theta-\theta_d)$ for the same
numerical data.}
    \label{resct}
\end{figure}

We note that the lifetime is super-Arrhenius. Fig. \ref{resct}
shows the logarithm of the lifetime, obtained by numerical
integration of eq. (\ref{velocity2}) and rescaled respectively by
$\tau_o$ (on the left) and by $\tau_o^{des}$ (on the right) as a
function of the energy barriers
$(\overline{\sigma_c}-\sigma_i)^2/2\theta$ (on the left) and
$(\overline{\sigma_c}-\sigma_i)^2/2(\theta-\theta_d)$ (on the
right). Eq. (\ref{lifetime}) does not rescale the data when
$\theta_d$ is varied. But, it is clear that the rescaling by eq.
(\ref{lifetime2}) makes all the data for different $\theta$ and
$\theta_d$ fall very well on the $y=x$ straight line. This
confirms the relevance of eq. (\ref{lifetime2}) which predicts an
increase of the lifetime with $\theta_d$. Once again, analytical
and numerical calculations confirm the deceleration of the crack
growth due to disorder.

\section{Facing the results with previous theoretical works}
The results presented in this letter differ from previous
theoretical works on one-dimensional Disordered Fiber Bundle Model
(1d-DFBM) describing the influence of disorder on the fracturing
process \cite{Ciliberto,Politi,Roux,Sornette}. In this model, as
long as the critical fraction of broken fibers is less than $50\%$
\cite{Sornette}, the introduction of a threshold distribution
actually accelerates rupture with an effective temperature larger
than the thermodynamical temperature. The rupture process starts
from the weaker fibers and progressively affects the stronger
ones. As a result, the distribution of breaking thresholds is
truncated from below by a front moving with time towards higher
thresholds. Because rupture can occur everywhere in the system,
this dynamics is non-local and the damage rate is given by how
many fibers break at each time step. Consequently, the effect of
disorder is obtained by averaging the damage rate over the
threshold distribution. This is not at all the process which
occurs when a large crack propagates in a disordered system. Here,
the probability of breaking at the crack tip is greatly enhanced
by stress concentration effects even for a large disorder. Then,
the rupture dynamics becomes local, and one needs to average over
disorder the time to break the fiber at the crack tip, i.e. the
inverse of the damage rate. Finally, it is worth noticing that
there are similarities between our model for crack growth
predicting a super-Arrhenius rupture time and the trap model
introduced by Bouchaud \cite{Bouchaud} to describe dynamics in
glassy systems.

\section{Discussion on the straight line growth hypothesis}
In our analysis, we assumed the crack to grow on a straight line
even if roughness is experimentally observed for sub-critical
cracks in heterogeneous materials. Is this hypothesis relevant for
crack growth in heterogeneous materials? Two viewpoints can be
raised.

One could think that roughness is due to the fact a crack, at each
moment, grow in the direction where the material is the weakest.
In this case, the crack, during its growth, will not experience
the intrinsic threshold distribution of the material, but only a
part of this distribution that corresponds mainly to the lowest
thresholds. Then, the distribution to be used in eq.
(\ref{distribution}) is not the intrinsic threshold distribution
of the material.

But, one could also think that crack growth occurs in the
direction where the stress intensification is the largest.
Statistically, the larger the stress, the higher the probability
to grow in the corresponding direction. This idea is supported by
recent theoretical and experimental works \cite{Procaccia1,
Procaccia2}. In this framework where the crack tip stress
distribution rules the roughness, it is fair to consider the
intrinsic threshold distribution of the material to rule the crack
growth.

Experimental reality is probably a compromise between the two
competing mechanisms: growth governed by stress intensification
and growth following the path through the weakest regions. In the
case of a material with a mesoscopic structure, such as paper, the
crack tip is surrounded by a finite number of fibers so that the
crack can grow only in a finite number of directions. The
determination of the growth direction is clearly due the
competition between stress intensification on each fiber compared
to its toughness. It is worth noticing that crack tip structure
complexity can create a complex stress distribution. Consequently,
the largest stress is not necessarily in the crack main direction
so that roughness can appear even if the crack always grows in the
maximum stress direction.

\begin{figure}[h]
\psfrag{L}[l]{$\ell_i$} \onefigure[width=13cm]{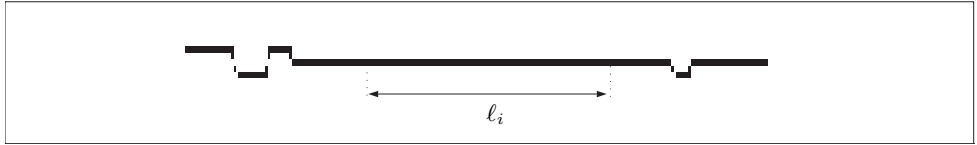}
\caption{Image of a crack created from an initial crack of length
$\ell_i$ by simulating thermal activation in the hexagonal spring
lattice.} \label{growthpattern}
\end{figure}

Some numerical work has been performed to illustrate a situation
where the crack is allowed to "choose" between a finite number of
growth directions. We model a two-dimensional elastic system as a
network of springs forming a lattice. More details about the
simulations are given in \cite{Santucci1}. The only discrepancies
with the simulations presented there are the use of a hexagonal
lattice instead of the previously used square lattice and the
introduction of a rupture threshold distribution.
\begin{figure}[h] \psfrag{X}[l]{$\sigma$}
\onefigure[width=5cm]{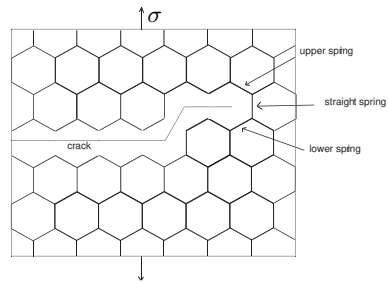} \caption{Geometry of the
hexagonal spring lattice of the numerical simulation with a
non-straight crack.} \label{hexa}
\end{figure}
We can see in fig. \ref{hexa}, that the crack can actually
"choose" between three directions (so called upper, straight and
lower directions) at each step. In fig. \ref{growthpattern}, a
typical crack pattern after a thermally activated growth is
presented. We can notice that the crack grows almost straight.
Actually, the stress intensification on the upper and lower
springs is only about $80\%$ of the one of the straight spring.
So, the crack will grow through the springs on the sides only if
the rupture threshold of the straight spring is very large. This
is only a rare event and we can say that the crack experiences
essentially the intrinsic threshold distribution truncated of the
very large thresholds only. In this simulation, the crack path is
mainly ruled by stress intensification.
Experimentally, the number of fibers joining at the crack tip can
be larger than three so that variations in stress intensification
between fibers are probably smoother than in the simulation.

No definitive conclusion on the growth process can be given
because, depending on the considered material, one of the two
competing mechanisms (growth governed by stress intensification
and growth following the path through the weakest fibers) will
dominate. The results presented in this letter hold essentially
for media where the crack path is ruled by stress intensification.

\section{Conclusion}

Assuming the existence of thermal stress fluctuations in the
material, we modelled sub-critical crack growth in disordered
materials. The influence of material disorder on the growth of a
single crack in two-dimensional brittle elastic material is
described through the introduction of a rupture threshold
distribution. Analytical predictions of the crack velocity and
material lifetime have been derived in agreement with direct
numerical calculations. The conclusion is that the crack growth
process is inhibited by disorder: velocity decreases and lifetime
increases with the temperature of disorder $\theta_d$. The
influence of disorder is simply accounted for by introducing an
effective temperature $\theta_{\text{eff}}=\theta-\theta_d$ for
thermodynamical temperatures above $\theta_d$. Hence, lifetime
follows a super-Arrhenius law. More experimental work is needed to
confirm this super-Arrhenius regime. In addition, the analogy with
glassy systems suggests that the growth of a crack in a disordered
material could also occur in a "glassy" regime, i.e. when
$\theta<\theta_d$.

\acknowledgments

We thank J.-L. Barrat for insightful discussions.

\end{document}